\documentclass[published]{JHEP3}
 \accepted{August 16, 2002}
 \received{July 31, 2002}
 \JHEP{08(2002)027}

\usepackage{epsfig} 

\newcommand{\be}[3]{\begin{equation}  \label{#1#2#3}}
\newcommand{\ee}{\end{equation}}
\newcommand{\ba}{\begin{array}}
\newcommand{\ea}{\end{array}}

\newsavebox{\uuunit}
\sbox{\uuunit}
    {\setlength{\unitlength}{0.825em}
     \begin{picture}(0.6,0.7)
        \thinlines
        \put(0,0){\line(1,0){0.5}}
        \put(0.15,0){\line(0,1){0.7}}
        \put(0.35,0){\line(0,1){0.8}}
       \multiput(0.3,0.8)(-0.04,-0.02){10}{\rule{0.5pt}{0.5pt}}
     \end {picture}}
\newcommand{\unity}{\mathord{\!\usebox{\uuunit}}}

\newcommand\U{\mathop{\rm {}U}}
\newcommand\SU{\mathop{\rm SU}}
\newcommand\SO{\mathop{\rm SO}}
\newcommand\SL{\mathop{\rm SL}}
\newcommand\Sp{\mathop{\rm Sp}}
\newcommand\Spin{\mathop{\rm Spin}} 

\title{Intersecting 6-branes from new 7-manifolds with $G_2$ holonomy}

\author{Klaus Behrndt\\
Max-Plank-Institut f\"ur Gravitationsphysik, 
Albert Einstein Institut\\ 
Am M\"uhlenberg 1,  14476 Golm, Germany\\
E-mail: \email{behrndt@aei.mpg.de}}

\author{Gianguido Dall'Agata and Dieter L\"ust\\
Humboldt Universit\"at zu Berlin,
Institut f\"ur Physik,\\ 
Invalidenstrasse 110, 10115 Berlin, Germany\\
E-mail: \email{dallagat@physik.hu-berlin.de}, 
\email{luest@physik.hu-berlin.de}}

\author{Swapna Mahapatra\\
Physics Department, Utkal University\\
Bhubaneswar 751 004, India\\
E-mail: \email{swapna@iopb.res.in}}

\abstract{We discuss a new family of metrics of 7-manifolds with $G_2$
holonomy, which are ${\mathbb R}^3$ bundles over a quaternionic space.
The metrics depend on five parameters and have two abelian
isometries. Certain singularities of the $G_2$ manifolds are related
to fixed points of these isometries; there are two combinations of
Killing vectors that possess co-dimension four fixed points which
yield upon compactification only intersecting D6-branes if one also
identifies two parameters. Two of the remaining parameters are
quantized and we argue that they are related to the number of
D6-branes, which appear in three stacks.  We perform explicitly the
reduction to the type IIA model.}

\begin{document}

\section{Introduction}

It is well known that the compactification of M-theory (11-dimensional
supergravity) on seven-manifolds $M_7$ of $G_2$ holonomy leads to an
effective theory in four dimensions with ${\cal N}=1$
supersymmetry. If $M_7$ is smooth, the harmonic Kaluza-Klein
decomposition of the 11-dimensional massless degrees of freedom leads
in four dimensions to ${\cal N}=1$ supergravity coupled to abelian
vector multiplets plus chiral multiplets, which correspond to the
moduli of $M_7$~\cite{820,370}. On the other hand, if $M_7$ exhibits
some singularities at certain points in the moduli space, massless
non-abelian gauge bosons possibly together with massless chiral matter
fields may emerge. The local neighborhood of these types of
singularities can be best described by replacing the compact space
$M_7$ by a non-compact $G_2$ manifold $X_7$ and we are essentially
dealing with the geometric description of the effective low-energy
gauge theory in four-dimensions (geometric engineering of gauge
theories).  In the following we are interested in M-theory on a
non-compact background $X_7$ for which a number of examples have been
discussed recently e.g.\ in~\cite{650}--\cite{860}.

If $X_7$ has a suitable $\U(1)$ isometry, one obtains a type IIA
superstring interpretation upon dimensional reduction to ten
dimensions. This circle is usually non-trivially fibred over a
six-dimensional base $B_6$ which serves as the geometric background of
the corresponding IIA superstring theory.  In order to obtain
non-abelian gauge groups with possibly chiral matter additional
D6-branes have to wrap supersymmetric 3-cycles of $B_6$.\pagebreak[3]
As a consequence, the gauge bosons correspond to open strings on the
D6-brane world volumes, and chiral fermions arise from open strings
stretching between different intersecting D6-branes.  In this way,
intersecting brane world models with intersecting D6-branes, being
wrapped on homology 3-cycles of 6-dimensional tori, orbifolds or
Calabi-Yau three-folds, can be constructed, which are more or less
closely related to the standard
model~\cite{Blumenhagen:2000wh}--\cite{840} (see e.g.~\cite{840} for a
more complete list of references on intersecting brane world models).
In M-theory language non-abelian gauge bosons arise, if $X_7$ has an
A-D-E singularity of codimension four. The non-abelian gauge bosons
correspond to massless M2-branes wrapped around collapsing 2-cycles
and product gauge groups are provided by intersecting singularities.
Massless fermions are supported by isolated (conical) singularities of
codimension 7 of $X_7$ and this situation can be realized by two or
more A-D-E singularities colliding into each other.  In the IIA brane
picture this is described by the intersection of D6-branes.  One can
also consider orientifold O6-planes (O6-planes correspond to the
Atiyah-Hitchin manifold) intersected by D6-planes. E.g. an O6-plane
intersected by $n$ D6-branes plus their mirror branes can lead to a
$\SU(n)$ gauge theory with chiral matter in the antisymmetric
represenation of $\SU(n)$.  In M-theory this corresponds to unfold a
$D_n$ singularity into a $A_{n-1}$ singularity.

Of course the IIA description depends very much on the choice of the
$\U(1)$ action. In order to obtain a configuration that contains only
D6-branes, one has to ensure that the 7-manifold has only co-dimension
4 fixed points and no co-dimension 2 and 6 fixed point sets.  In this
case, the 6-branes could be embedded in a topologically flat space and
following the arguments given in~\cite{550,740,150} the topology of
the 7-manifold should be completely encoded in the fixed point set of
the $\U(1)$ action.  In this case we can expect to describe a known
4-dimensional field theory living on the common intersection.

So far not many explicit metrics are known. Basically they group
together into two classes~\cite{090,080}: one is topologically a
$\mathbb R^4$ bundle over $\mathbb S^3$ and the other a $\mathbb R^3$
bundle over a quaternionic base space.  Many generalizations, with
more parameters or functions, have been discussed in the past years.
The first class e.g., can be generalized to $\mathbb R^4/\mathbb Z_N$
bundle over $\mathbb S^3$. In the second class one can consider
further quaternionic spaces, different from e.g.\ the 4-sphere
${\mathbb S}^4$ and the complex projective space ${\mathbb C \mathbb
P}^2 = {\SU(3) / \U(2)}$, which are the only compact homogeneous
quaternionic 4-dimensional spaces~\cite{320}.  Apart from their
non-compact analogs, there are also non-homogeneous quaternionic
spaces as discussed in~\cite{730,860,881,880}. For a closely
related discussion of quaternionic spaces appearing in hyper Kaehler
cones see~\cite{990,991,992}.

In this paper we want to discuss a $G_2$ metric based on a
quaternionic space with only two isometries. This 4-dimensional
Einstein manifold can be obtained by a Wick rotation of a solution
found by Demianski and Plebanski~\cite{020,010} and is given by four
roots of a fourth order polynomial. After some general comments about
manifolds with $G_2$ and $\Spin(7)$ holonomy in the next section, we
will discuss the quaternionic space and its symmetries in
section~\ref{section3}. In section~\ref{section4} we will discuss in
detail the fixed point set of the two Killing vectors.  Following the
standard lore~\cite{360,740,150}, we identify 6-branes as co-dimension
four fixed points and avoid co-dimension two and six fixed point
sets. Finally, in section~\ref{section5} we perform the dimensional
reduction and obtain explicit forms of the type IIA fields.

\section{Manifolds with $G_2$ and $\Spin(7)$ holonomy from 
quaternionic spaces}

Consider M-theory on the manifold $M_4 \times X_7$ where $M_4$ is the
flat 4-d Minkowski space. The resulting 4-d field theory exhibits
${\cal N}$=1 supersymmetry if $X_7$ allows for exactly one
(covariantly constant) Killing spinor and in the absence of
$G$-fluxes this is the case if the manifold $X_7$ has $G_2$ holonomy.
The exceptional group $G_2$ appears as automorphism group of
octonions: $o = x^0 {\mathbb I} + x^a i_a$, where $i_a$ satisfy the
algebra
\[
i_a i_b = -\delta_{ab} + \psi_{abc} \, i_c \,,
\]
and the $G_2$-invariant 3-index tensor $\psi_{abc}$ is given in the
standard basis by
\begin{eqnarray} 
\Psi &=& {1 \over 3!} \psi_{abc} \, e^a \wedge e^b \wedge e^c 
\nonumber\\
&=& e^1 \wedge e^2 \wedge e^3 + e^4 \wedge e^3 \wedge e^5 +
e^5 \wedge e^1 \wedge e^6 + e^6 \wedge e^2 \wedge e^4 + 
\nonumber\\ &&{}
+e^4 \wedge e^7 \wedge e^1 +e^5 \wedge e^7 \wedge e^2 +
e^6 \wedge e^7 \wedge e^3 \,, 
\nonumber\\ 
&=& e^1 \wedge e^2 \wedge e^3 + {1 \over 2} \, 
	e^i \wedge e^m \wedge J^i_{mn} e^n
\label{100}
\end{eqnarray}
where $J^i_{mn}$ ($i = 1,2,3$, $m=4,5,6,7$) are the anti-selfdual
($J^i_{mn} = -{1 \over 2} \epsilon_{mnpq} \, J^i_{pq}$) complex
structures defined by the algebra
\be110
J^i \cdot J^j = - \unity \, \delta^{ij} + \epsilon^{ijk} J^k \,.
\ee
$G_2$-holonomy requires that this 3-index tensor is closed and co-closed 
\be120
d \Psi = d^{\star}\Psi=0 
\ee
which implies that $\Psi$ is a covariantly constant 3-form and is
equivalent to the existence of a Killing spinor.  This in turn is
ensured if the spin connection satisfies the projector
condition~\cite{110, 100}
\be130
\psi_{abc} \, \hat \omega^{bc} = 0 \,.
\ee
Both conditions~(\ref{120}) and~(\ref{130}) yield a set of first
order differential equations for the metric functions. If the manifold
allows for more covariantly constant form-fields, the holonomy is
further restricted and the Killing spinor equation has more than one
solution so that the 4-dimensional model has extended supersymmetry.

As it has been shown in~\cite{090, 080} (see also~\cite{730} where
our notations are used) a metric that fulfills these equations
is given by
\be140
ds^2 = {1 \over \sqrt{2 \kappa |u|^2 + u_0}}\, 
	\big( du^i + \epsilon^{ijk} A^j u^k \big)^2
	+ \sqrt{2 \kappa |u|^2 + u_0} \; ds^2_4 \,.
\ee
which is topologically a ${\mathbb R}^3$ bundle (related to the
coordinates $u^i$) over a quaternionic base space, given by the metric
$ds^2_4$ with the curvature $\kappa$ and the $\SU(2)$ connection $A^i$
($u_0$ is an integration constant); see next section for our
conventions. This $G_2$ metric is, up to $\SU(2)$ rotations of the
complex structures, fixed by the quaternionic base space and in the
next section we discuss in detail the quaternionic space that we want
to consider. In the limit $\kappa = 0$ this space becomes
hyper-K\"ahler with vanishing $\SU(2)$ curvature and hence the
connection $A^i$ gives a pure gauge transformation, see
eqs.~(\ref{220}) and~(\ref{240}). Therefore, the connection part
in~(\ref{140}) can be absorbed by a proper $\SU(2)$ rotation of the
$u^i$ coordinates and the space becomes a direct product of $\mathbb
R^3$ and the hyper K\"ahler space. But also if the curvature is
non-trivial, there is still the freedom to choose a proper $\SU(2)$
basis.

For $\kappa \neq 0$ we can also introduce polar coordinates for the
${\mathbb R}^3$ part and the metric becomes
\be150
ds^2 = {dr^2 \over \kappa (1 - {4u_0 / r^4})} + {r^2 \over 4 \kappa}
\left(1 - {4u_0 \over \, r^4}\right) 
g_{ab} \left(dx^a + \xi^a_i A^i\right) \left(dx^b + \xi^b_j A^j\right)
+ {r^2 \over 2} ds^2_4\,,
\ee
where $g_{ab}$ is the metric of $\mathbb S^2$ with the three Killing
vectors $\xi^a_i$. In the limit $u_0 \rightarrow 0$ this metric is a
cone over a 6-manifold $Y$ which is a $\mathbb S^2$ bundle over the
quaternionic space $Q$ and this manifold has a weak $\SU(3)$
holonomy. To see this we write the 7-metric (with $u_0 =0$ and for
$\kappa = 1 $) as
\begin{equation}
ds^2 = dr^2 + r^2 ds^2_{Y} \,.
\end{equation}
Decomposing the fibered ${\mathbb R}^3$ as
\begin{eqnarray}  
     u^1 &=& |u|\cos \theta\,,\nonumber\\ 
     u^2 &=& |u| \sin \theta \,\cos \varphi\,,\nonumber\\ 
     u^3 &=& |u| \sin \theta \, \sin \varphi\, ,\label{281}
\end{eqnarray}
the metric of the six-dimensional base becomes
\begin{equation}
ds^2_{Y}= V^a \otimes V^a\,,
\end{equation}
where
\begin{eqnarray} 
\label{V6}
V^1 &=& \frac {\hat{e}^1}{r} \equiv \frac12 \left(d \theta - \sin
\varphi A^2 + \cos \varphi A^3\right),
\nonumber\\ 
V^2 &=& \frac {\hat{e}^2}{r} \equiv \frac12 \left(\sin \theta\,d
\varphi + \sin \theta A^1-\cos\theta\cos\varphi A^2 - \cos \theta
\sin\varphi A^3\right),
\nonumber\\ 
V^{m} &=& \sqrt{\kappa\over 2} \, e^m_4
\end{eqnarray}
(where $e^m_4$ is the vielbein of the quaternionic space).  We can now
show that this manifold is half-flat, which, according to~\cite{760},
implies a reduction to $\SU(3)$ defined by $\omega$ and $\psi_\pm$ for
which $\hat{d} \psi_+ = 0$ and $\omega \wedge \hat{d} \omega = 0$, but
$\hat{d} \omega \neq 0$ (where the differential $\hat{d}$ is taken on
the six-dimensional subspace).  This implies that $Y$ has weak
$\SU(3)$ holonomy, as it is expected.  {From} the $\SU(3)$ forms one can
build the harmonic 3-form $\Psi$ which defines the $G_2$ structure~as
\begin{equation}
\Psi = \omega \wedge dr + \psi_+\,.
\end{equation}
In our case the two-form $\omega$ is given by
\begin{equation}
\omega \equiv \hat{e}^1 \hat{e}^2+ \hat{e}^3 \hat{e}^4+\hat{e}^5
\hat{e}^6\,,
\end{equation}
and the three-form $\psi_{+}$ satisfies
\begin{equation}
\psi_+ \equiv  \frac13 d \omega =
\hat{e}^1\hat{e}^3\hat{e}^5-\hat{e}^1\hat{e}^4\hat{e}^6-
\hat{e}^2\hat{e}^3\hat{e}^6 
- \hat{e}^2\hat{e}^4\hat{e}^5\,.
\end{equation}
The $\SU(3)$ reduction is completed by another three-form $\psi_{-}$,
defined such that they satisfy the compatibility relations $\omega
\wedge \psi_{\pm} = 0$ and $\psi_+ \wedge \psi_- = \frac23 \omega^3$.
We have already explicitly constructed $\hat{e}^1$ and $\hat{e}^2$
in~(\ref{V6}) and we can obtain the rest of the six-dimensional
orthonormal base $\hat{e}^i$ performing a $\theta$ and $\varphi$
dependent $\SO(4)$ rotation of the seven-dimensional base~$e^i$:
\begin{eqnarray}
r \hat{e}^3 &=& \sin \theta \, e^4 + \cos\theta \left( \cos \varphi \,
e^5 + \sin \varphi \,e^6\right),
\\
r \hat{e}^4 &=& \cos \varphi \,
e^6 - \sin \varphi \,e^5\,,
\\
r \hat{e}^5 &=& - e^7\,,
\\
r \hat{e}^6 &=& -\cos \theta \, e^4 + \sin\theta \left( \cos \varphi \,
e^5 + \sin \varphi \,e^6\right).
\end{eqnarray}

Let us end this section with a comment on 8-manifolds with $\Spin(7)$
holonomy. Again, they allow for one (covariantly constant) Killing
spinor and yield therefore ${\cal N}$=1 supersymmetry in three
dimension upon dimensional reduction.  The construction is again fixed
by a 4-d quaternionic space $Q$ and the metric reads~\cite{090, 080}
(see also~\cite{691, 740} for generalizations)
\be160 
ds^2 = {dr^2 \over \kappa (1- {u_0 / r^{10/3}})} + {9 \over 100 \,
\kappa} r^2 \left(1 - {u_0 \over r^{10/3}}\right) \left(\sigma^i -
A^i\right)^2 + {9 \over 20} \, r^2 \, ds^2_4
\ee
where $u_0$ is again an integration constant and $\sigma^i$ are
the left-invariant one-forms on $\SU(2)$. Topologically, this
space is an $\mathbb R^4$ bundle over the quaternionic space
and the cone $Y$ (orbits of constant $r$) is now an $\mathbb S^3$
bundle over $Q$.

\section{Quaternionic space with two commuting isometries}\label{section3}

In the last section we have introduced the class of manifolds with
$G_2$ and $\Spin(7)$ holonomy, which are basically fixed by a
quaternionic base space. In this section we will consider a specific
quaternionic space with two isometries that we later-on want to employ
for $G_2$ spaces.

\subsection{General conventions}

Quaternionic-K\"ahler spaces are complex spaces that allow for three
complex structures $J^i$ ($i = 1,2,3$) defined by the
algebra~(\ref{110}). Denoting the quaternionic vielbein by $e^m$, one
obtains three 2-forms $\Omega^i$ by
\be210
\Omega^i = -\frac{\kappa}{2} \, e^m \wedge J^i_{mn} e^n \,.
\ee
The holonomy of a 4n-dimensional quaternionic spaces is contained in
$\Sp(n) \times \SU(2)$. This statement is trivial for $n=1$ and can be
replaced by the requirement that the Weyl-tensor of 4-dimensional
quaternionic space has to be anti-selfdual
\[
W + {\, ^{\star}W} = 0 \,.
\]
For a quaternionic space in any dimension the triplet of 2-forms
$\Omega^i$ is expressed in terms of the $\SU(2)$-part of the
quaternionic connection $A^i$ as
\be220
dA^i + {1 \over 2} \, \epsilon^{ijk} A^j \wedge A^k = 
 \Omega^i 
\ee
which ensures that the triplet of 2-forms is covariantly constant.
Moreover, any quaternionic space is an Einstein space with 
curvature $\kappa$ implying that its metric $g_{mn}$ solves the
equation
\be230
R_{mn} = 3 \, \kappa \, g_{mn} \,.
\ee
The complex structures can be selfdual or anti-selfdual and in our
notation we will take the latter ($J^i_{mn} = -{1 \over 2}
\epsilon_{mnpq} \, J^i_{pq}$) so that the triplet of 2-forms can be
written as
\begin{eqnarray} 
\Omega^1 &=&- \kappa \left( e^4 \wedge e^7 - e^5 \wedge e^6\right),
\nonumber\\ 
\Omega^2 &=& -\kappa \left(e^4 \wedge e^6 + e^5 \wedge e^7 \right),
\nonumber\\ 
\Omega^3 &=& -\kappa \left(-e^4 \wedge e^5 + e^6 \wedge e^7\right).
\label{240}
\end{eqnarray}
Moreover, the $\SU(2)$ connection is given as the anti-selfdual part
of the spin connection $\omega^{mn}$ of the quaternionic space
\be250
A^i = {1 \over 2} \omega^{mn} J_{mn}^i \,.
\ee
In the same way, the selfdual part gives the $\Sp(n)$ connection.

\subsection{Deriving the explicit metric}

The maximally symmetric 4d quaternionic space has 10 isometries
spanning a group of rank two ($\SO(5)$ or $\SO(4,1)$) and hence there
are at most two commuting isometries.  We are interested in the
situation, where the space admits only these two isometries and all
others are broken. This can be done by a double orbifold, which
imposes non-trivial periodicities along these two directions.  Hence,
consider the metric ansatz
\be260
ds^2_4 = {1 \over F^2(p,q)} 
	\left[ {dp^2 \over P(p)} + P(p) \, d\tau^2
	+ {dq^2 \over Q(q)} + Q(q) \, d\sigma^2 \right]
\ee
where $\partial_\tau$ and $\partial_\sigma$ are the two commuting
Killing vectors and (single) zeros of $P$ and $Q$ require non-trivial
periodicity in $\tau$ and $\sigma$. Since the metric has to be
Einstein, we can derive the function $F(p,q)$ from the combination of
the Ricci tensor
\[
0= R_p^{\ p} - R_\tau^{\ \tau} = 2 F \, \partial_p^2 F \,,\qquad
0= R_q^{\ q} - R_\sigma^{\ \sigma} = 2 F  \partial_q^2 F \,.
\]
Taking as solution $F = p+q$ and calculating another combination of
the Ricci tensor yields
\[ 
0=\partial_p \partial_q \left( {R_\sigma^{\ \sigma} - R_\tau^{\ \tau}
	\over p + q }\right) = {1 \over 2} \left[ Q'''(q) - P'''(p)
	\right]
\]
and therefore $P$ and $Q$ are polynomials of third degree. It is
straightforward to investigate the other equations and one finds as
general solution of the equation~(\ref{230}): $P = a_0 - \kappa + a_1
p + a_2 p^2 + a_3 p^3$, $Q = -a_0 + a_1 q - a_2 q^2 + a_3 q^3$.  The
Weyl tensor for this space is anti-selfdual only if: $a_3 = 0$.  So,
this quaternionic space depends in total on four parameters that fix
the identifications for $\sigma$ and $\tau$. The torus spanned by
these two isometries is diagonal, but one can also deform the torus
while keeping the quaternionic property.  Fortunately, the
corresponding metric has been known for quite some time. It was
introduced as Minkowskean solution by Demianski and
Plebanski~\cite{020,010} and a discussion in the mathematical
literature is given e.g.\ in~\cite{050,040,030}, see
also~\cite{060,510,881,880} for more general quaternionic spaces with
two isometries.  The corresponding euclidean metric reads
\be270
ds^2_4 = {1 \over (1 + pq)^2} \!\left[{ p^2 - q^2 \over P} dp^2 + {p^2 -
	q^2 \over Q}dq^2 + {P \over p^2 -q^2} \left(d \tau + q^2 d
	\sigma \right)^2 {+} {Q \over p^2 - q^2} \left(d \tau + p^2 d
	\sigma \right)^2 \right]
\ee
where the polynomials are now given by $P = \alpha - 2n p - \epsilon
p^2 + 2m p^3 + (\alpha - \kappa) p^4$, $Q = - \alpha + 2m q + \epsilon
q^2 - 2n q^3 - (\alpha - \kappa) q^4$ and the Weyl tensor becomes
anti-selfdual iff: $m=n$. Again (single) zeros of $P$ and $Q$ are conical
singularities, which gives periodicities of $\sigma$ and $\tau$
defining the deformed torus. In order to recover the form~(\ref{260}),
one makes the transformation $p \rightarrow 1/p$ combined with $(p, q,
\tau, \sigma) \rightarrow {1 \over \lambda} (p, q, \tau, \sigma)$ and
$(\alpha, n, \epsilon, m, \kappa) \rightarrow (\alpha, \lambda n,
\lambda^2 \epsilon, \kappa)$ followed by the limit $\lambda
\rightarrow \infty$, see also~\cite{020, 010}.

However, we do not want to use this form of the metric and apply
another scaling: $(p,q) \rightarrow {\lambda} (p,q)$, $(\alpha, n,
\epsilon, m, \kappa) \rightarrow ({\alpha \lambda^4}, {n \lambda^3},
{\epsilon \lambda^2}, {m \lambda^3}, \kappa)$ followed  by the
limit $\lambda \rightarrow 0$. As a consequence the metric becomes
\be280 
ds^2_4 = { p^2 - q^2 \over P} dp^2 + {p^2 - q^2 \over Q}dq^2 + {P
\over p^2 -q^2} \left(d \tau + q^2 d \sigma \right)^2 + {Q \over p^2 -
q^2} \left(d \tau + p^2 d \sigma \right)^2
\ee
with
\be290
P = \alpha - 2n p - \epsilon p^2 - \kappa p^4 \,, \qquad
Q = - \alpha + 2m q + \epsilon q^2 + \kappa q^4  \,.
\ee
In the following we will use this form of the metric, which has
again an anti-selfdual Weyl tensor iff: $m=n$. In this case
the two polynomials become, up to the overall sign, identical and we 
can use the notation
\begin{eqnarray} 
P &=& - \kappa (p-r_1)(p-r_2)(p-r_3)(p-r_4) \,,\nonumber\\ 
Q &=&  \kappa (q-r_1)(q-r_2)(q-r_3)(q-r_4) \,,\nonumber\\ 
0 &=& r_1 + r_2 +r_3 +r_4 \,.
\label{300}
\end{eqnarray}
In addition to the two abelian isometries, there are the following
symmetries
\begin{eqnarray} 
(i) && p \leftrightarrow q \,,\nonumber\\ 
(ii) && p \rightarrow -p \,, \qquad q \rightarrow -q \,, \qquad
	r_i \rightarrow -r_i \,, \nonumber\\ 
(iii) && (p, q, \tau, \sigma) \rightarrow \left(\lambda \, p, \lambda
	\, q, \frac {\tau}{\lambda}, \frac
	{\sigma}{\lambda^3}\right)\qquad {\rm and}\quad r_i
	\rightarrow \lambda \, r_i \,.
\label{302}
\end{eqnarray}
The last symmetry can be used to scale one non-vanishing parameter to
$\pm 1$. We have therefore the following interpretation of the
parameters: one is obviously the cosmological constant, two
parameterize the orbifolds and turning off one of them yields an
\pagebreak[3] enhancement of one $\U(1)$ isometry to one of three
different groups $\SU(2)$, $\SL(2,R)$ or the Heisenberg group, that
are related to the three discrete values of the fourth parameter (see
also next subsection).

It is important to note, that the physical parameter range is given by
the values of $(p,q)$ which fulfill the two inequalities
\be320 
(p^2 - q^2) \, P(p) \geq 0 \qquad {\rm and} \quad (p^2 - q^2)
\, Q(q) \geq 0 \,.
\ee
This allows for a number of different coordinate regions, which are
separated by regions that contain two timelike coordinates.  Note,
these timelike regions appear beyond fixed points of the isometries,
which become branes upon dimensional reduction. Thus, they indicate
the appearance of additional massless modes and should be interpreted
as phase transition points.

An important property of this space is the presence of a curvature
singularity, which becomes visible in the square of the Riemann
curvature
\be310
R_{abcd} R^{abcd} = {24 \kappa^2 } +  {96 \, n^2 \over (p + q)^6} 
\ee
where $n$ was the coefficient of the linear part in the polynomials.
This co-dimension one singularity at $p + q =0$ is present for any
value of the fiber coordinates $u^i$ and hence is a singularity also
of the 7-manifold (actually, it is singular domain wall of the whole
11-dimensional space time). There are two limits in which this
curvature singularity disappears. One is obviously given by $n=0$ and
the other by $n \rightarrow \infty$ combined with a proper rescaling
of $p$ and $q$. As we will discuss in the next section, both limits
yield a homogeneous quaternionic space; $\mathbb S^4$ or
$\mathbb{CP}^2$ (or their non-compact versions).

Having the metric it is straightforward to determine the 
$\SU(2)$ connections as introduced in~(\ref{250}). They are given by
\begin{eqnarray} 
A^1 &=&  {\sqrt{PQ} \over (q-p)} \, d \sigma \,,
\nonumber\\ 
A^2 &=& { -\kappa(p-q)} \, d\tau + {1 \over (p-q)}
	\left[\alpha - n (p+q) - \epsilon \, qp - \kappa p^2 q^2 \right] 
	\, d \sigma
\nonumber\\ 
A^3 &=& {1 \over (p-q)} \left[ \sqrt{Q \over P} \, dp +
	\sqrt{P \over Q} \, dq \right].
\end{eqnarray}
and fulfill the relations~(\ref{220}) and~(\ref{240}) with
\begin{eqnarray} 
e^4 &=&  \sqrt{p^2 - q^2} \frac{dp}{\sqrt{P}} \,, \qquad\qquad\qquad
e^5 =  -\sqrt{p^2 - q^2} \frac{dq}{\sqrt{Q}}\,, 
\nonumber\\ 
e^6 &=&  \frac{\sqrt{P}}{\sqrt{p^2 - q^2}} \left( d\tau + q^2 
d\sigma\right),\qquad 
e^7 =  -\frac{\sqrt{Q} }{\sqrt{p^2 - q^2}} \left(d\tau + p^2 
d\sigma\right).
\end{eqnarray}

\subsection{Special limits}

The 4-dimensional base space as introduced in the last subsection, can
be obtained by a Wick rotation of a solution that has been discussed
by Plebanski and Demianski as a ``Rotating, Charged, and Uniformly
Accerating Mass in General Relativity''~\cite{110}.  It is
\pagebreak[3] also known
as the (A)dS-Kerr-Newman-Taub-NUT solution, where the electric and
magnetic charges are obviously zero in our application.  To make the
relation to these known Einstein spaces more clear, let us perform the
corresponding limits.

To obtain the euclidean (A)dS-Kerr-Newman-Taub-NUT solution as a limit
of our euclidean PD solution, we set (see also~\cite{750})
\begin{eqnarray} 
q &=& r\,, \qquad p = a\cos\theta + N\,, \qquad 
\tau = t +  \left (\frac{N^2}{a} + a \right )\frac{\phi}{\Xi}\,, 
\qquad 
\sigma = - \frac{\phi}{a\Xi}\,, 
\nonumber\\ 
\alpha &=& - a^2 + N^2 \left (1 - \kappa \, 3a^2 + \kappa \, 3N^2
\right ), 
\nonumber\\ 
n &=& N \left [ 1 - \kappa \, a^2 + 4 \kappa N^2 \right ], 
\nonumber\\ 
\epsilon &=& -1 - \kappa \, a^2 - 6 \kappa N^2\,, 
\nonumber\\ 
\Xi &=& 1 - \kappa a^2 \,.
\label{394}
\end{eqnarray}
With these transformations and relaxing the constraint $m=n$ (so that
the Weyl tensor is not anti-selfdual), the polynomials $P$ and $Q$
become
\begin{eqnarray}
P &=& - a^2\sin^2\theta \left [ 1 - \kappa
\left ( 4 a N \cos\theta + a^2 \cos^2\theta \right ) \right ], 
\\ 
Q &=& - (r^2 + N^2) + \kappa \left (
r^4 - a^2r^2 - 6 N^2 r^2 + 3 a^2 N^2 - 3 N^4 \right ) + 2 m r + a^2 \,.
\end{eqnarray}
If we moreover define, 
\begin{eqnarray} 
R^2 & = & r^2 - (a\cos\theta + N)^2  \,, 
\\
\lambda & = & (r^2 + N^2) - \kappa
\left ( r^4 - a^2 r^2 - 6 N^2 r^2 + 3 a^2 N^2 - 3 N^4 \right ) 
- 2 m r - a^2  
\end{eqnarray}
one gets, 
\begin{eqnarray}
\frac{p^2 - q^2}{Q(q)} dq^2  & = & \frac{R^2}{\lambda} dr^2 \,, 
\nonumber\\
\frac{p^2 - q^2}{ P(p)} dp^2 & = &  \frac{R^2}
{1 - \kappa (a^2\cos^2\theta + 4 a N \cos\theta)} d\theta^2 \,, 
\nonumber \\
\frac{Q(q)}{p^2 - q^2} \left ( d\tau + p^2 d\sigma \right )^2
& = &  
\frac{\lambda}{R^2} \left [ dt + \frac{a \sin^2\theta - 2 N\cos\theta}
{\Xi} d\phi \right ]^2 
\nonumber \\
\frac{P(p)}{p^2 - q^2} \left (d\tau + q^2 d\sigma 
\right )^2  & = &
\frac{\sin^2\theta \left [ 1 - \kappa a \cos\theta (a \cos\theta + 
4 N)\right ] }{R^2}   
\left [ a dt - \frac{(r^2 - a^2 - N^2)}{\Xi} d\phi \right ]^2  
\nonumber
\end{eqnarray}
and we obtain the euclidean (A)dS Kerr-Taub-NUT solution given by
\begin{eqnarray}
ds^2_4 & = & \frac{R^2}{1 - \kappa (a^2\cos^2\theta + 
4 a N \cos\theta)}
d \theta^2 + \frac{R^2}{\lambda} dr^2 + 
\nonumber \\
&&{}+ \frac{\lambda}{R^2} \left [ dt + \left ( \frac{a \sin^2\theta}
{\Xi} - \frac{2 N\cos\theta}{\Xi} \right ) d\phi \right ]^2 + 
\nonumber \\
&& {}+\frac{\sin^2\theta \left [ 1 - \kappa (a^2\cos^2\theta + 4 a N
    \cos\theta)  
\right ]} {R^2} \left [ a dt - \frac{(r^2 - a^2 - N^2)}{\Xi} 
d\phi \right ]^2 .
\end{eqnarray}
The limits are now straightforward: if $N=0$ one obtains the euclidean
(A)dS-Kerr solution, where $a$ corresponds to the rotational
parameter.  But note, there is no rotation in an euclidean space, the
axial symmetric minkowskean Kerr-solution becomes instead
\pagebreak[3] an euclidean dipole solution. In fact, the euclidean
Kerr solution (i.e.\ for $\kappa = 0$) has been identified
in~\cite{790} as Taub-NUT/anti-Taub-NUT dipole solution where the
parameter $a$ just measures the distance between the two centers.  On
the other hand, if $a=0$ while $N \neq 0$, $\kappa \neq 0$, the
solution becomes euclidean (A)dS-Taub-NUT given by~\cite{450}
\begin{eqnarray}
ds^2_4 = V(r) \left ( dt - 2 N\cos\theta d\phi \right)^2 + 
\frac{dr^2 }{V(r)}  + (r^2 - N^2) \left ( d\theta^2 + 
\sin^2\theta d\phi^2 \right ) 
\end{eqnarray}
where
\begin{equation} 
V(r) \equiv \frac{\lambda}{R^2} = \frac{1}{r^2 - N^2}
\left [ (r^2 + N^2) - \kappa \left ( r^4 - 6 N^2 r^2 - 
3 N^4 \right ) - 2 m r \right ].
\end{equation} 
The isometry group for this space has been enhanced to $\U(1) \times
\SU(2)$ (from $\U(1) \times \U(1)$ for $a \neq 0$) and the relevance of
this quaternionic space in gauge supergravity and for $G_2$ manifolds
has been discussed recently in~\cite{120,730}.  In the limit of
vanishing $N$, the space becomes euclidean $(A)dS_4$ (i.e.\ $\mathbb
S^4$ or the non-compact hyperboloid), which is maximal symmetric with
10 isometries parameterizing $\SO(5)$ or $\SO(4,1)$.  On the other hand,
in the limit $N\rightarrow \infty$ while keeping $\hat r = N( r - N)$
fix, the solution becomes the coset space ${\SU(3) / \U(2)}$
($=\mathbb{CP}^2$) or ${\SU(2,1) / \U(2)}$ resp. This is the second
known regular 4-dimensional quaternionic space, which has 8 isometries
parameterizing $\SU(3)$ or $\SU(2,1)$.  It is also instructive to
understand these limits in terms of the four roots $r_m$ as introduced
in~(\ref{300}). The maximal symmetric spaces ($\mathbb S^4$ resp.\
$EAdS_4$) can be obtained if
\be232
r_1= - r_4 \,, \qquad r_2 = -r_3
\ee
and the corresponding transformation is given in~\cite{110} (for
Minkowskean signature). On the other hand, for $N \rightarrow
\infty$ we find from~(\ref{394}): $\alpha = 3 \kappa N^4$, $n = 4
\kappa N^3$, $\epsilon = - 6\kappa N^2$ yielding: $P(p) = -\kappa (-3
N^4 + 8 N^3 p - 6 N^2 p^2 + p^4) = \kappa (N-p)^3(p+3N)$. Thus,
one gets $\mathbb{CP}^2$ or its non-compact analog in the limit where
three zeros of the polynomial coincide, as e.g.\ 
\be382
r_2=r_3=r_4=N \qquad {\rm and} \quad N \rightarrow \infty \,.
\ee
This limit is of course only regular if one shifts also $q$ and $p$
(see eqs.~(\ref{394}) and recall the replacement $r = {\hat
r / N} + N$, see also~\cite{120,730}).

\section{Fixed point set}\label{section4}

\subsection{General discussion}

The quaternionic space has two Killing vectors and let us consider the
isometry obtained by the linear combination
\begin{equation}
k = \beta_1 \partial_\tau - \beta_2 \partial_\sigma\, .
\end{equation}
Since the $\SU(2)$ connection $A^i$ does not depend on $\sigma$ and
$\tau$, this Killing vector corresponds to an isometry also of the
$G_2$ manifold~(\ref{150}).  To find the fixed points of such
isometry, we have to satisfy the equations $|k|^2 = 0$ and with the
metric~(\ref{150}) we find
\begin{eqnarray}  \label{k2}
|k|^2 & =& {r^2 \over 4 \kappa} \left(1 - {4 u_0 \over
 r^4}\right)  g_{ab} \xi^a_i \xi^b_i \, (\beta_1 A^i_\tau - \beta_2
A^i_\sigma)^2 + 
\nonumber\\  &&{}
+ {r^2 \over 2} \left[ \frac{P}{p^2 - q^2}
  \left(\beta_1 - \beta_2 q^2\right)^2+ \frac{Q}{p^2 - q^2}
  \left(\beta_1 - \beta_2 p^2\right)^2 \right] = 0\,.
\end{eqnarray}
This is one necessary condition on fixed points, but in order to
ensure that it is at finite geodesic distance one has to require that
the fixed point set is non-degenerate, i.e.\ $(\nabla k)^2 \neq 0$ at
$|k|^2 = 0$. If this condition is not fulfilled the space exhibits a
infinite throat and the fixed point will be at infinity (gauging such
an isometry in gauged supergravity results in a run-away solution,
see~\cite{120}). Actually in this case the Killing vector does not
parameterize a rotational symmetry, but a translational one.  This
happens e.g., if the fourth order polynomials have a double zero, see
also the discussion below.

Since the physical parameter range of the $(p,q)$ coordinates is given 
by the values which fulfill the inequalities
\begin{equation}
\label{830}
\left(p^2-q^2\right) P(p) \geq 0 \qquad \hbox{ and } \quad 
\left(p^2-q^2\right) Q(q) \geq 0 \,,
\end{equation}
each term in~(\ref{k2}) has to vanish separately. Apart from the
trivial zero at $r = 0$, the second term of $|k|^2$ vanishes for the
two cases
\begin{eqnarray} 
\label{solu}
(a)&&   P = 0 \qquad \hbox{and} \quad Q = 0 \,, 
\nonumber\\ 
(b)&& \ p =  \sqrt{\beta_1 \over \beta_2} = r_m \qquad \hbox{or}
\quad q =  \sqrt{\beta_1 \over \beta_2} = r_m 
\end{eqnarray} 
where $r_m$ is one of the roots of $P(p)$ [respectively $Q(q)$].  The
condition $(a)$ fixes $p$ and $q$ at points where the two isometric
$\U(1)$ fibers in the metric vanish and hence this condition defines a
point on the quaternionic space and is called a NUT.  On the other
hand, condition $(b)$ fixes only one coordinate ($p$ or $q$) and only
one $\U(1)$ fiber vanishes and therefore this condition defines a
2-dimensional subspace --- a bolt. Obviously this latter case can only
happen for a specific Killing vector, a generic choice of $\beta_1$
and $\beta_2$ will not yield bolts. For both cases $(a)$ and $(b)$
only $A^2_\sigma$ and $A^2_\tau$ are non-trivial and hence the first
term in~(\ref{k2}) is zero iff
\begin{eqnarray} 
(c) \qquad |\xi_2|^2 &=& 0 \,, \nonumber\\ 
(d) \quad\  A^2_\mu \, k^\mu& =& 
	\beta_1 A^2_{\tau} - \beta_2 A^2_\sigma  = 0 \quad {\rm or}
	\nonumber\\ 
(e)\qquad\quad  r^4 &=& 4 u_0 \,.
\label{918}
\end{eqnarray}
The last case is only of interest as long as $u_0 \neq 0$ and
corresponds to the point where the $\mathbb S^2$ has collapsed to a
point while the quaternionic space is still finite.  Case $(c)$ is
satisfied at fixed points of the second $\mathbb S^2$-Killing vector
(i.e.\ $|\xi_2|^2 = 0$) and this gives exactly two (antipodal) points
on $\mathbb S^2$, which in the coordinates~(\ref{281}) are given by
$\cos\theta = \sin\varphi = 0$ (or $u^1 = u^3 =0$). For case $(d)$ one
finds
\be379
\beta_1 A_\tau^2 - \beta_2 A^2_\sigma
	= { (\beta_1 - \beta_2 q^2) \big[ (p-q) \partial_p - 2 \big] P - 
	(\beta_1 - \beta_2 p^2)\big[(p-q) \partial_q  + 2 \big]Q  
	\over 2(p-q)^2(p+q)} 
\ee
and this has to vanish in combination with case $(a)$ or $(b)$.  By
inserting the polynomials~(\ref{300}) one finds, for generic values of
$\beta_1$ and $\beta_2$, that this can only happen at double zeros of
$P$ or $Q$.  But as we discussed before these double zeros correspond
to degenerate fixed points which are not at finite geodesic distance.
On the other hand, in combination with case $(a)$, we find always a
ratio of ${\beta_1 \over \beta_2} = \beta$ for which this combination
vanishes at zeros of $Q$ and $P$, see also the explicit example
below. Notice, for these simple Killing vectors the
combination~(\ref{379}) gives the Killing prepotential (momentum maps)
and for a 4-dimensional quaternionic space with at least two abelian
compact Killing vectors there is exactly one combination for which the
Killing prepotentials (or momentum maps) vanish at the fixed
point~\cite{900, 500}.

In summary, depending on the choice of parameters there are fixed
point sets of various co-dimensions:
\begin{description}
\item[Fixed point set of co-dimension 7]

These are zeros of $|k|^2$ which are points on the 7-manifold.  This
is the case at the conical singularity at $r=0$ if $u_0=0$ or otherwise
a combination of the constraint $(a)$ with $(e)$.

\item[Fixed point set of co-dimension 6]

They are related to a combination of case $(a)$ and $(c)$, which means
that the fixed point set is given by a NUT on the quaternionic space
combined with a fixed point of the second $\mathbb S^2$ Killing
vector. Since we have two abelian isometries we can first reduce over
the $k$ to get a IIA configuration followed by a T-duality over the
second isometry. In this procedure, these co-dimension 6 fixed points
should be mapped onto type IIB NS5-branes, because they are fixed
points of both isometries of the 7-manifold and hence are also fixed
points of translations along the T-duality direction.

\item[Fixed point set of co-dimension 5]

They are only present if $u_0 \neq 0$ and correspond to a combination
of case $(b)$ and $(e)$, but they are not additional isolated fixed
point sets.  In fact, $r^4= 4 u_0$ represents exactly the point of
minimal distance of given codimension 4 fixed points set.  The same is
true for the codimension 7 fixed point appearing as combination of
case $(a)$ and $(e)$, which is the orbit of minimal distance between
given co-dimension 6 fixed point sets.

\item[Fixed point set of co-dimension 4]

These are perhaps the most interesting ones, since they are identified
as 6-branes upon the reduction to type IIA string theory.  We obtain
co-dimension 4 fixed points as a combination of case $(b)$ and $(c)$
as well as of case $(a)$ and $(d)$. In both situations the 6-branes
will wrap a 2-cycle of the 6-manifold $Y$: for the combination $(b)$
and $(c)$ this 2-cycle is the bolt inside the quaternionic space and
if $p$ and $q$ are bounded by two roots $r_m$, this 2-cycle is
topologically a line segment times a circle and if there are no
conical singularities this 2-cycle becomes topologically an $\mathbb
S^2$.  Recall, case $(b)$ as well as case $(d)$ require 
specific Killing vectors which \emph{do not} agree, but in any
case 6-branes appear only for a proper choice of the $11th$
coordinate.  For the combination $(a)$ and $(d)$, the 6-branes are
transversal to the quaternionic space and wrap all three $u^i$
coordinates.

\item[Fixed point set of co-dimension 2]

\looseness=-1 These fixed points can appear only as a combination of
case $(b)$ and case $(d)$. As we mentioned after equation~(\ref{379})
this requires that $p$ or $q$ run toward a double zero of $P$ or $Q$
and from the metric~(\ref{280}) we see that these double zeros are not
at finite geodesic distance. Instead, near these points the space
develops a throat and the fixed point set is at infinity and we can
discard them.  Alternatively, case $(d)$ can appear for a specific
Killing vector, which is however different from the one fixed by case
$(b)$ and hence there are no co-dimension 2 fixed points.
\end{description}

Recall, in addition to these fixed points  there is the co-dimension one
curvature singularity of the quaternionic space at $p+q = 0$.

\subsection{Nuts and bolts of the quaternionic space}

In order to determine the number of fixed points, we have to ask for
the number of solutions of the equations~(\ref{solu}), which are
related to NUTs and bolts on the quaternionic space. Given that our
polynomial $P$ or $Q$ has four roots $r_m$ one can distinguish among
four main cases:
\be251
\ba[b]{ll}
i) & \kappa >0 \  \mbox{while} \ r_2, r_3, r_4 \geq 0\  \mbox{and} \
r_1 <0 \,, 
\\[2mm]
ii) &\kappa >0\  \mbox{while}\  r_3, r_4 \geq 0 \,, \ r_1,r_2 \leq 0\ 
	\mbox{and}\  r_4+r_1 >0\,,\\[2mm]
iii) &\kappa <0\  \mbox{while} \ r_2, r_3, r_4 \geq 0\  \mbox{and} \ 
	r_1 < 0 \,,\\[2mm]
iv) & \kappa <0\  \mbox{while} \ r_3, r_4 \geq 0\,, \ r_1,r_2 \leq 0\  
	\mbox{and}\   r_4+r_1 \geq 0 \,.
\ea
\ee
Every other case can be reconducted to one of the above upon using
some of the symmetries~(\ref{302}) of the metric~(\ref{280}). 

Let us start with the discussion of the possible bolts.  Any of the
four roots for which $\sqrt{\beta_1 / \beta_2} = r_m$ gives bolts
and since $p$ and $q$ can go independently to this root there are
always two bolts. But note, not each coordinate region contains a
bolt.  E.g.\ if $p$ is in region IV (see figure) and $q$ in region III
we have two bolts if $r_m = \sqrt{\beta_1 / \beta_2} = r_3$. On
the other hand, if $\sqrt{\beta_1 / \beta_2} = r_1$ one finds
bolts only if one takes into account the other allowed coordinate
regions, namely that $p$ is in region II and $q$ in region III or vice
versa.

The discussion of all possible NUTs is more involved.  It can be shown
that one finds six solutions (less if some equality bounds are
satisfied or if there are double roots) for any of the above
possibilities in~(\ref{251}).  They are never grouped in more than
three in the same connected physical region of parameters.  Actually
one can find the following patterns: zero, one or three fixed points
if the region does not contain the $p=-q$ singularity, two fixed
points when the region contains the $p=-q$ singularity.  We give here
a table summarizing such possibilities, where we grouped the fixed
points according to the $(p,q)$ sector they belong. For the cases with
$\kappa >0$ we find the fixed points summarized in table~\ref{tab1}
whereas $\kappa <0$ gives the fixed points summarized in table~\ref{tab2}

{\renewcommand{\belowcaptionskip}{12pt} 
\TABLE[t]{\begin{tabular}{|c|c|c|c|}
    \hline
$(i) $ & $(r_4,r_1) $ & $ (r_2,r_3)$ , $(r_2,r_4)$ , $(r_3,r_4) $ &
$(r_2,r_1)$ , $(r_3,r_1)$ \\[1mm]
\hline
$(ii)$ & $(r_3,r_4)$ , $(r_2,r_4)$ & $(r_2,r_1)$ ,
$(r_3,r_1)$ ,
$(r_3,r_2)$ &
$(r_1,r_4)$ \\[1mm]
    \hline 
\end{tabular}\caption{These are all values of $(p,q)$ that are NUT fixed
points of the quaternionic space with the parameters
defined in (\ref{251}). The fixed points in the same group
are in the same physical region of parameters.}\label{tab1}}

\TABLE[t]{\begin{tabular}{|c|c|c|c|c|c|}
    \hline
$(iii)$ &
$(r_4,r_3)$ &
$(r_4,r_2)$ &
$(r_3,r_2)$ &
$(r_1,r_3)$ ,
$(r_1,r_4)$ &
$(r_1,r_2)$   \\[1mm] \hline
$(iv)$ &
$(r_4,r_3)$ &
$(r_4,r_2)$ ,
$(r_4,r_1)$ &
$(r_2,r_3)$ &
$(r_1,r_3)$ &
$(r_1,r_2)$   \\[1mm] \hline
\end{tabular}\caption{These are the analogous fixed points for a negatively
curved quaternionic space.}\label{tab2}}}

Note, in the degenerate case where two or three roots are equal, one
looses physical regions, which were defined by the
relations~(\ref{830}) and recall, at double zeros the space develops a
throat and effectively cuts the space in two disconnected regions. On
the other hand, if $p$ and $q$ approach a single zeros from opposite
sites this point is regular and one can pass this point.

\FIGURE[t]{\centerline{\epsfig{file=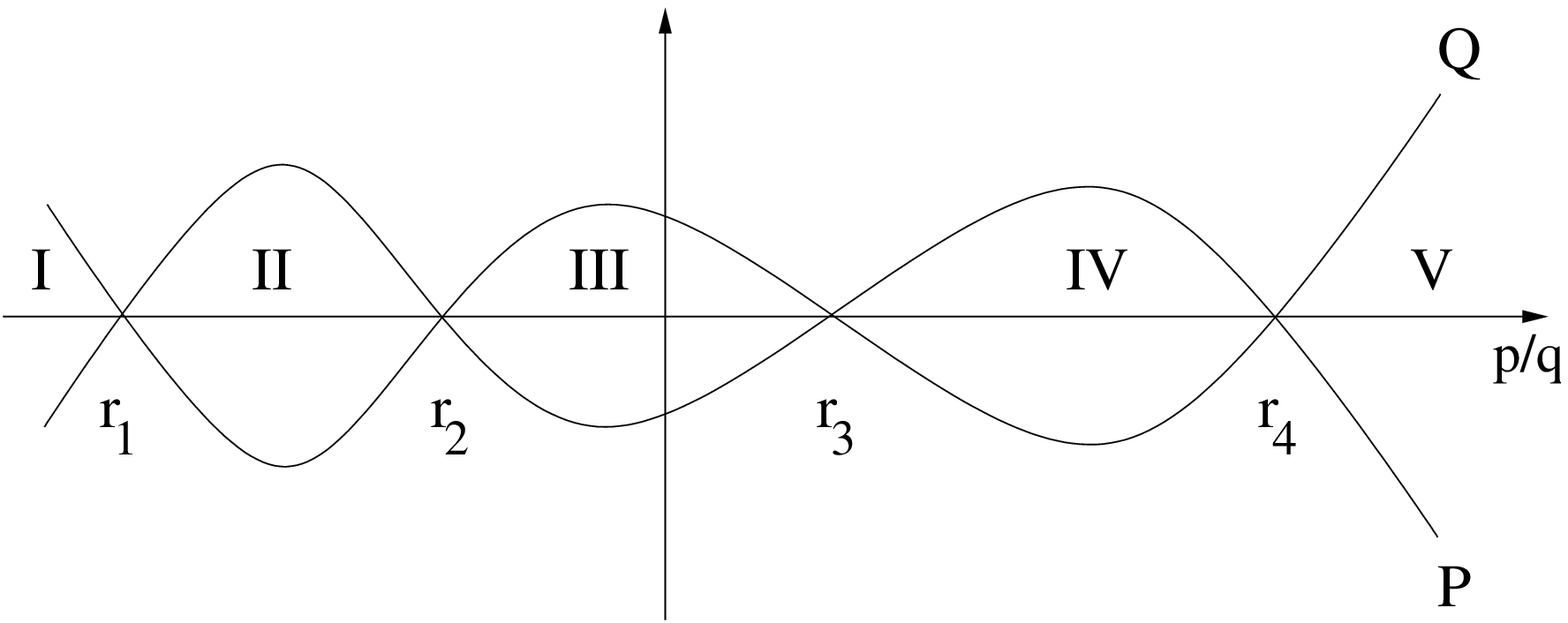, width=100mm}\caption{The
quaternionic space is basically defined by two fourth order polynomial
$P(p)$ and $Q(q)$ which differ only by a total sign. In this figure we
have shown the case $\kappa > 0$ and denoted with $I, II, \ldots , V$
the different coordinate regions. The case with negative $\kappa$
corresponds effectively to an exchange of $P$ and $Q$.}\label{fig1}}}

\subsection{Explicit example}

Now we want to describe an explicit example which has only
co-dimension 4 fixed points that become D6-branes upon
compactification.  Since there will be no other fixed points, the
number of D6-branes is related to the number of co-dimension 4 fixed
points~\cite{360,150,740}. As for the $\mathbb{CP}^2$ case, we will
find that the fixed point set has two components and hence there are
in total three stacks of D6-branes.  An interesting question is to
determine the number of 6-branes located at each fixed point. For the
standard 6-brane, this number is related to the periodicity in the
Taub-NUT space that resolves the conical singularity and also here,
the number of 6-branes should be related to the conical singularity
appearing at the fixed point.  The corresponding deficit angle is
given by the surface gravity of the corresponding fixed point
set. This is a well-known quantity discussed in black hole physics,
which is defined by $(\nabla k)^2_i$ calculated on the fixed point set
$\Gamma_i$.  It can be shown that this quantity is constant over the
fixed point set and it gives, multiplied with the area of the fixed
point set, the contribution to the Noether charge related to the
Killing symmetry, see~\cite{800,810}. Upon compactification this
Noether charge gives the D6-brane charge.  Applied to black holes, the
surface gravity is the Hawking temperature, which is nothing but the
inverse periodicity that resolves the conical singularity. It is
straightforward to show, that the Taub-NUT space with NUT charge $N$
has a surface gravity of $|\nabla k| \sim {1 / N}$ and therefore
we will identify the number of 6-branes of the fixed point set
$\Gamma_i$ by: $N_i \sim {1 / |\nabla k|_i}$.

\pagebreak[3]

Let us now come to the concrete example. Recall, in order to obtain
co-dimension 4 fixed points one has to consider a particular Killing
vector so that
{\renewcommand{\theenumi}{\roman{enumi}}  
\begin{enumerate} 
\item condition $(b)$ in eq.~(\ref{solu}) is satisfied 
or alternatively 

\item the expression $(d)$ in~(\ref{918}) vanishes at a zero of $P$
and $Q$.
\end{enumerate}}
\noindent A closer look on equation~(\ref{379}) shows that both cases
can only happen at the same time at double zeros of the polynomial
yielding degenerate fixed points. We will therefore consider both
cases independently.

For case $(i)$ we consider the Killing vector
\be771
k =r_3^2 \, \partial_\tau - \partial_\sigma 
\ee
which means that $r_3^2 = \beta_1$ ($\beta_2 =1$) and without further
restrictions we will assume that $r_3 > 0$.  In this example we will
consider $p$ in region IV and $q$ in III or vice versa (see
figure). There are now two sets of 6-branes located at
\be221
\ba[b]{ll}
D6_1 :\quad &  p = r_3 \,,\qquad u_1 = u_3 = 0 \,, \\[1mm]
D6_2 :\quad &  q = r_3 \,,\qquad u_1 = u_3 = 0 \,.
\ea
\ee
But by keeping generic values of the roots, there will be further
codimension 6 fixed points at $q=r_2$, $p=r_4$, $u_1=u_2=0$ and at
$p=r_2$, $q=r_4$, $u_1=u_2=0$. In order to avoid these fixed points we
will set $r_1 = r_2$, which essentially moves these fixed points to
infinity since the metric~(\ref{280}) develops an infinite throat at
$p \rightarrow r_2=r_1$.  Calculating the surface gravity for the
fixed point set given in eq.~(\ref{221}) gives
\begin{equation} 
|\nabla k|_{1} =|\nabla k|_2 = {\kappa \over 4} \, (3 r_3+r_4)^2
 (r_4-r_3)
\end{equation} 
where we used the constraint $0=r_1+r_2+r_3+r_4 = 2r_2+ r_3 + r_4$.
That both numbers coincide, is a consequence of the symmetry $p
\leftrightarrow q$ of the metric.

Next, let us consider the 6-branes coming from case $(ii)$, where the
Killing vector
\be779
k = \beta \, \partial_\tau - \partial_\sigma
\ee
was fixed so that eq.~(\ref{379}) vanishes at a
zero of $P$ and $Q$.  Recall, this corresponds to the $\U(1)$ isometry
for which the Killing prepotentials vanish, see~\cite{900,500}. To be
concrete, we will consider the fixed point: $p=r_4$ and $q=r_3$ and
find that~(\ref{379}) vanishes at this point if
\be882
\beta = {\beta_1 \over \beta_2} = {1 \over 2} \left[ r_3(r_2 - r_4)
	+ r_2 (r_2 + r_4) \right].
\ee
This yields 6-branes located at
\be229
\ba{ll}
D6_1 :\quad &  p = r_4 \,,\qquad q= r_3\,, \\[1mm]
D6_2 :\quad &  p = r_3 \,,\qquad q = r_4 \,.
\ea
\ee
Both are NUTs on the quaternionic space and all three $u^i$
coordinates are part of their worldvolume. In order to avoid
additional co-dimension 6 fixed points if $p$ (or $q$) run toward
$r_2$ (see figure), we set again: $r_1 = r_2$ and $p$ (or $q$) becomes
a non-compact coordinate and can be identified as the overall radial
coordinate for the 6-brane intersection.  This time we find for the
surface gravity
\begin{equation} 
|\nabla k|_{1} =|\nabla k|_2 = \kappa 
\frac{r_3 r_4}{r_3 + r_4}(2 r_3 + r_4)(2 r_4 + r_3)\,. 
\end{equation} 
As we argued at the beginning, this surface gravity should be related
to the number D6-branes of two of the three stacks of 6-branes, which
should be a quantized number. As explained in~\cite{460,740}, a
consistent $\U(1)$ action requires that the ratio of two eigenvalues of
the 2-form $dk$ calculated at the fixed point set should be a rational
number. For the case at hand, this gives the quantization condition
\be827
{n \over m} = {1 + \sqrt{1 -\Delta} \over |\Delta|}\, , \qquad  
\Delta = {2 (r_3^2 - r_4^2) \over r_4^4 + r_3^4 + 2}
\ee
where $n$ and $m$ are relative prime integers. This condition ensures,
that a tangent vector at the fixed point comes back to its own if we
go once around the circle (the $\U(1)$ action of a Killing vector
acts as a rotation in the tangential plane given by two rotational
parameters for a co-dimension 4 fixed point set).

\section{Type IIA reduction}\label{section5}

We will discuss now the reduction of the previous example 
along the compact direction determined by the isometry: 
\begin{equation}
\label{dz}
\partial_z = \partial_\sigma-\beta \partial_\tau\,.
\end{equation}
This generic reduction becomes relevant to the intersecting 
$D6$-branes setup explained in the previous section when $\beta = 
r_{3}^2$ or it satisfies~(\ref{882}).
{From} this choice we can introduce two new coordinates $w$ and $z$ 
\begin{equation}
\sigma = z \,, \qquad \tau = -\beta \,z + \,w\,,
\end{equation}
such that $z$ becomes the coordinate along which we perform the
reduction, while $w$ completes the set of the remaining
ten-dimensional ones
\begin{equation}
x^\mu = \{ y^a, u^i, p, q,w\} \qquad ( a =0,\ldots3)\,, \quad(i=1,2,3)\,.
\end{equation}
Here $y^a$ parametrize ${\mathbb R}^{1,3}$, $u^i$ were introduced
before and parametrize ${\mathbb R}^3$ and $p$, $q$ and $w$ are the
surviving coordinates of the quaternionic part.  Performing the
reduction using the standard Kaluza-Klein ansatz followed by a
conformal rescaling, one produces a 10-dimensional IIA bosonic
background with non-vanishing dilaton $\phi$ and metric in the NS-NS
sector, whereas only the one-form $C_{\mu}$ is turned on in the RR
sector.  The relation between the two metrics, which also fixes the
dilaton dependence, is given by
\begin{equation}
ds^2_{(11)}= e^{-\frac23 \phi(x)} \, g_{\mu\nu} dx^\mu dx^\nu + 
e^{\frac43 \phi(x)} \,\left[dz + dx^\mu C_\mu(x)\right]^2.
\end{equation}

\pagebreak[3]

It is interesting to point out that the dilaton and the one-form can
be completely determined in terms of the Killing vector~(\ref{dz})
upon which we reduce the metric.  This is much simpler in this new
coordinate system.  The relations for the dilaton and the one-form~are
\begin{eqnarray}
\label{dilatonk}
e^{\frac43 \phi} &=& g_{zz} = |k|^2 \,,\\[5pt]
\label{Ck}
C &=& e^{-\frac43 \phi} g_{zM} dx^M = |k|^{-2} \, k_M dx^M \,.
\end{eqnarray}
A first consequence is that the component of the one-form along the 
Killing vector direction is fixed
\begin{equation}
{\imath}_{k}C \equiv C_M k^M = 1\,,
\end{equation}
which implies that $C$ has the right number of independent components.
As a further consequence, one can determine its field strength in
terms of $k$ and its derivatives
\begin{equation} \label{329}
F_{MN}= 2 \partial_{[M} C_{N]} = 6\, {k^S k_{[S} \nabla_M k_{N]} \over
|k|^4} \,.
\end{equation}
Since near a brane configuration we expect to have some flux in ten
dimensions, this should show up in the integral of the two-form $F$ on
the transversal two-cycle $\cal C$:
\begin{equation}
\int_{\cal C} F \neq 0\,.
\end{equation}
\looseness=1 From the relations above and the fact that this flux should be
quantized, we expect that its number could be read from the
eigenvalues of the $\nabla k$ matrix.  This is also compatible with
the picture given in the previous section, where the surface gravity
was related to the number of $D6$-branes and this latter was also
derived from the $dk$ two-form.

We can now proceed to give the explicit expression for the various
ten-dimensional fields.  To do so, it is useful to define the
following quantity
\begin{equation}
{\cal A} = {\imath}_k A^2 = 
\frac{\alpha - n (p+ q) - \epsilon \, p q 
-\kappa \, p^2 
q^2+\beta \kappa (p-q)^2}{p-q}\,,
\end{equation}
whose vanishing is related to the appearance of the co-dimension four 
singularities for the NUT fixed points of the quaternionic manifold.
As we will see, this quantity appears repeatedly in the following 
formulae and this let us simplify the structure of the dilaton and 
$C$-field equations.
The dilaton is determined to be
\begin{eqnarray} 
e^{\frac43 \phi} &=& \displaystyle\frac{1}{\sqrt{2 \kappa 
|u|^2 +u_0}} \left[ 2 u_1 \,
u_2\, \frac{\sqrt{PQ}}{(p-q)}\,{\cal A} + 
\frac{PQ}{(p-q)^2} \left( u_2^2 + u_3^2\right) 
+ {\cal A}^2\,\left(u_1^2 + 
u_3^2\right)\right] 
\nonumber\\[8pt]
&&{}+ \displaystyle\sqrt{2 \kappa 
|u|^2 +u_0} \,\frac{P\, \left(q^4 + \beta^{2} - 2\, \beta \,q^2\right)+ 
Q\,(p^4 + \beta^2 - 2 \,\beta \,p^2)}{p^2 - q^2}\,, 
\label{dilaton}
\pagebreak[3] \end{eqnarray} 
and the one-form $C$ is
\begin{eqnarray} 
C &=&  \frac{ e^{-\frac43 \phi} }{\sqrt{2 \kappa 
|u|^2 +u_0}} \biggl\{ \left[
{\cal A} (u_3\, du_1-u_1 \,du_3) + 
\frac{\sqrt{PQ} (u_3 \,du_2-u_2 \,du_3)}{(p-q)} \right]+
\nonumber\\ &&
\hphantom{\frac{ e^{-\frac43 \phi} }{\sqrt{2 \kappa 
|u|^2 +u_0}} \biggl\{}
+  u_3\left[  
\frac{(p-q) \,{\cal A}\, u_2-\sqrt{PQ} \, u_1}{(p-q)^2} \right] 
\left(\,\sqrt{\frac{Q}{P}}\, dp + \sqrt{\frac{P}{Q}} \,dq\right)\,+
\nonumber\\&&
\hphantom{\frac{ e^{-\frac43 \phi} }{\sqrt{2 \kappa 
|u|^2 +u_0}} \biggl\{}
+ dw  \left[-\kappa  (p{-}q)  {\cal A} (u_1^2 {+} u_3^2) 
+  \beta  \kappa  (p{-}q)^2 (u_1^2 {+} 
u_3^2)-\kappa  \sqrt{PQ} u_1 u_{2} \right] \biggr\}\!+
\nonumber\\ &&{}
+  e^{-\frac43 \phi} \, \sqrt{2 \kappa 
|u|^2 +u_0} \,\frac{P\,(q^2 - \beta)+ Q\,(p^2 - \beta)}{p^2-q^2} \,dw\,.
\end{eqnarray}
The reduced metric is then:
\begin{equation}
ds^2_{(10)}=e^{\frac23 \phi}\left[dx^a \, dx^b \,\eta_{ab}+ {1 \over
\sqrt{2 \kappa |u|^2 + u_0}}\left(du^i + \epsilon^{ijk} \tilde{A}^j
u^k\right)^2 + \sqrt{2 \kappa |u|^2 + u_0} \;d\tilde{s}_3\right]
\end{equation}
with
\begin{equation}
\tilde{A}^1= 0\,,\qquad 
\tilde{A}^2 = -\kappa (p-q) dw \,,\qquad
\tilde{A}^3 = {1 \over (p-q)} \left[ \sqrt{Q \over P} \,
dp + \sqrt{P \over Q} \, dq \right],
\end{equation}
and
\begin{equation}
d\tilde{s}_3 = { p^2 - q^2 \over P} dp^2 + {p^2 - q^2 \over Q}dq^2 +
	{{P+Q} \over p^2 -q^2} \,dw^2\,.
\end{equation}

From these explicit formulae we can now see that $w\to w+c$ is the
surviving $\U(1)$ isometry of the background, commuting with the
$\partial_{z}$ upon which we reduced the 11-dimensional solution.  One
can therefore think about the possibility of further reducing the
above solution to 9 dimensions along this direction and consider the
$T$-dual picture.  The interesting fact is that the corresponding
Killing vector does not show any fixed point (unless one considers a
double root of our $P$ and $Q$ polynomials, which then becomes an
essential singularity).  Let us conclude that in this reduction we do
not produce $NS 5$-branes in addition to the $D6$-brane setup
discussed in the previous section.

To be explicit, we consider now the above reduction in the case that
$\beta$ is chosen such that one can have codimension four
singularities at NUT fixed points of the quaternionic manifold, i.e.\
$\beta$ satisfying~(\ref{882}).  Doing this we expect to obtain a
setup of three intersecting $D6$-branes and we want to analyze the
behaviour of our solution when the coordinates approach the fixed
point corresponding to one of these branes.  The fixed point we will
discuss sits at
\begin{equation}
p = r_4\,, \qquad q = r_3\,,
\end{equation} 
and we choose to have $r_{1} = r_{2}$, such that the additional
codimension six singularities are removed.\footnote{The analysis for
the other fixed point is totally symmetric upon exchange of $p$ and
$q$.} The first thing to be pointed out is that at such fixed points
the string coupling constant vanishes, as the dilaton can be expressed
as the square of the Killing isometry, see~(\ref{dilatonk}), and this
latter must go to zero at the fixed points.  We can then proceed to
the analysis of the limit of the reduced 10-dimensional metric and
one-form.  Before proceeding with the limit, we have to remember that
the fixed point is found by fixing the value of two coordinates and
that therefore the limiting procedure has to be defined accordingly.
Since the surfaces $p=r_4$ or $q = r_3$ already show an irregular
behaviour for the $C$ field and the metric, we decided to approach the
fixed point in a ``diagonal" direction.  This means that we took a
similar scaling for $p$ and $q$, namely $p = r_4 - x$, $q = r_3-x$ and
then took the limit $x\to 0^+$ for $\kappa >0$.  In this way it can be
checked that the dilaton behaviour is linear in $x$
\begin{equation}
\label{dilx}
e^{\frac43 \phi} \simeq \frac{\left( 3\,r_3^2 +10\, r_3\, r_4 +
3\,r_4^2 \right)^2\, \left( 5\,r_3^2 +6\, r_3\,r_4 +5\,r_4^2\right) }
{128\,\left( r_3 +r_4 \right) }\, \, \sqrt{u_0 +2\,\kappa
\,|u|^2}\;\kappa\, x + O(x^2)\,,
\end{equation} 
and all the above quantities are positive.  The same limit in the
metric shows the expected behaviour for a $D6$-brane geometry, taking
care of the fact that in our parametrization the internal and
transverse space are not expressed through cartesian coordinates.  As
a matter of fact, it can be shown that the leading behaviour is given
by
\begin{equation}
g_{pp} \sim g_{qq} \sim \frac{1}{\sqrt{x}}\,,
\end{equation}
whereas
\begin{equation}
g_{ww} \sim g_{w.v.} \sim \sqrt{x}\,.
\end{equation}
Here we called $g_{w.v.}$ the world-volume metric and $p$, $q$ and $w$
are the transverse coordinates.  The behaviour of the metric in the
$w$ direction is different from the standard one shown by $p$ and $q$
because $w$ is an angular coordinate parametrizing the $\U(1)$
isometry of the resulting metric and therefore one has to add further
scaling coming from the radial direction.  Again, as expected, the
two-form field strength $F$ shows a diverging behaviour in the $p$ and
$q$ directions
\begin{equation}
F_{p\mu} \sim F_{q\mu}\sim \frac{1}{x}\,,
\end{equation}
whereas all the other components go to some constant value.  In line
of principle one could now also derive the exact number of $D6$-branes
sitting at such fixed point by integrating the $F$-form along the
collapsing two-cycle of the metric.  Unfortunately, as already shown
by the dilaton expression~(\ref{dilx}), the definitions of $F$ are
highly complex in our coordinate system and therefore we decided not
to perform such computation.

\section{Conclusion}

In the paper, we discussed in detail the metric of a new 7-manifold
with $G_2$ holonomy. This space is topologically a $\mathbb R^3$
bundle over a quaternionic space with a $\U(1) \times \U(1)$ isometry
group and is determined by a single fourth order polynomial. A generic
Killing vector has fixed points of various co-dimension, but most
interesting are co-dimension 4 fixed points that give D6-branes upon
dimensional reduction. As we discussed in detail, this requires to
pick specific Killing vectors and we found exactly two possibilities
to obtain D6-branes. In order to avoid additional co-dimension 6 fixed
points one has to equalize two roots of the fourth order polynomial.
The co-dimension four fixed point set consist of two components and we
concluded therefore that there are three stacks of D6-branes, where
two of the stacks have equal number of branes.

Following the arguments given in the mathematical
literature~\cite{030}, it is very tempting to relate this space to the
weighted projective space. In fact, the four roots of the fourth order
polynomial sum up to zero and hence are parameterized by three
(quantized) parameters, which should be related to the three weights
of $\mathbb{WCP}^2_{abc}$. In order to avoid co-dimension 6 fixed
points we had to identify two roots and the remaining two parameters
where quantized.  As a consequence, the number of 6-branes in two
stacks agree and we expect a gauge group $\SU(m) \times \SU(m) \times
\SU(n)$, where in the deformed case the higgsing should be done in
such a way that the product of two equal gauge groups survives;
because the two components of the fixed point set are related to the
same number of 6-branes.  At the moment, these conclusions are more
speculative and further investigations are necessary.

\acknowledgments

We would like to thank Andreas Brandhuber and Sergei Gukov for
valuable discussions. The work of K.B.\ is supported by a Heisenberg
grant of the DFG and G. D. acknowledges the financial support provided
through the European Community's Human Potential Programme under
contract HPRN-CT-2000-00131 quantum spacetime.  S.M.\ would like to
thank Alexander von Humboldt Foundation for financial support during
the initial stage of this work. S.M.\ also acknowledges the facilities
provided by the computer centre, I.O.P. Bhubaneswar.<


\begin{thebibliography}{10}


\bibitem{820}
M.J. Duff, B.E.W. Nilsson and C.N. Pope, \emph{Kaluza-Klein
  supergravity}, \prep{130}{1986}{1}.

\bibitem{370}
G.~Papadopoulos and P.K. Townsend, \emph{Compactification of $D = 11$
  supergravity on spaces of exceptional holonomy},
  \plb{357}{1995}{300} [\hepth{9506150}].

\bibitem{650}
B.S. Acharya, \emph{On realising $N=1$ super Yang-Mills in M-theory},
  \hepth{0011089}.

\bibitem{550}
M.~Atiyah, J.M. Maldacena and C.~Vafa, \emph{An M-theory flop as a
  large-$N$ duality}, \jmp{42}{2001}{3209} [\hepth{0011256}].

\bibitem{780}
M.~Cveti\v{c}, G.W. Gibbons, H.~Lu and C.N. Pope, \emph{M3-branes,
  $G_2$ manifolds and pseudo-supersymmetry}, \npb{620}{2002}{3}
  [\hepth{0106026}].

\bibitem{770}
A.~Brandhuber, J.~Gomis, S.S. Gubser and S.~Gukov, \emph{Gauge theory
at large-$N$ and new $G_2$ holonomy metrics}, \npb{611}{2001}{179}
[\hepth{0106034}].

\bibitem{360}
M.~Atiyah and E.~Witten, \emph{M-theory dynamics on a manifold of
  $G_2$ holonomy}, \hepth{0107177}.

\bibitem{140}
M.~Cveti\v{c}, G.W. Gibbons, H.~Lu and C.N. Pope, \emph{Cohomogeneity
  one manifolds of ${\rm Spin}(7)$ and $G_2$ holonomy},
  \prd{65}{2002}{106004} [\hepth{0108245}].

\bibitem{740}
S.~Gukov and J.~Sparks, \emph{M-theory on ${\rm Spin}(7)$ manifolds 1},
\npb{625}{2002}{3} [\hepth{0109025}].

\bibitem{Witten:2001uq}
E.~Witten, \emph{Anomaly cancellation on $G_2$manifolds},
  \hepth{0108165}.

\bibitem{350}
B.~Acharya and E.~Witten, \emph{Chiral fermions from manifolds of
  $G_2$ holonomy}, \hepth{0109152}.

\bibitem{Curio:2001dz}
G.~Curio, B.~K\"ors and D.~L\"ust, \emph{Fluxes and branes in type-II
  vacua and M-theory geometry with $G_2$ and ${\rm Spin}(7)$ holonomy},
  \npb{636}{2002}{197} [\hepth{0111165}].

\bibitem{Brandhuber:2001kq}
A.~Brandhuber, \emph{$G_2$ holonomy spaces from invariant
three-forms}, \npb{629}{2002}{393} [\hepth{0112113}].

\bibitem{150}
S.~Gukov and D.~Tong, \emph{D-brane probes of special holonomy
  manifolds and dynamics of $N=1$ three-dimensional gauge theories},
  \jhep{04}{2002}{050} [\hepth{0202126}].

\bibitem{730}
K.~Behrndt, \emph{Singular 7-manifolds with $G_2$ holonomy and
intersecting 6-branes}, \npb{635}{2002}{158} [\hepth{0204061}].

\bibitem{881}
C.I. Lazaroiu and L.~Anguelova, \emph{M-theory compactifications on
  certain `toric' cones of $G_2$ holonomy}, \hepth{0204249}.

\bibitem{880}
L.~Anguelova and C.I. Lazaroiu, \emph{M-theory on `toric' $G_2$ cones
  and its type-II reduction}, \hepth{0205070}.

\bibitem{870}
P.~Berglund and A.~Brandhuber, \emph{Matter from $G_2$ manifolds},
  \hepth{0205184}.

\bibitem{860}
M.~Cveti\v{c}, G.W. Gibbons, H.~Lu and C.N. Pope, \emph{Bianchi IX
  self-dual Einstein metrics and singular $G_2$ manifolds},
  \hepth{0206151}.

\bibitem{Blumenhagen:2000wh}
R.~Blumenhagen, L.~G\"orlich, B.~K\"ors and D.~L\"ust,
  \emph{Noncommutative compactifications of type-I strings on tori
  with magnetic background flux}, \jhep{10}{2000}{006}
  [\hepth{0007024}].

\bibitem{Aldazabal:2000cn}
G.~Aldazabal, S.~Franco, L.E. Ib\'a\~nez, R.~Rabad\'an and
A.M. Uranga, \emph{Intersecting brane worlds}, \jhep{02}{2001}{047}
[\hepph{0011132}].

\bibitem{Aldazabal:2000dg}
G.~Aldazabal, S.~Franco, L.E. Ib\'a\~nez, R.~Rabad\'an and
  A.M. Uranga, \emph{$D = 4$ chiral string compactifications from
  intersecting branes}, \jmp{42}{2001}{3103} [\hepth{0011073}].

\bibitem{Blumenhagen:2000ea}
R.~Blumenhagen, B.~K\"ors and D.~L\"ust, \emph{Type I strings with F-
  and B-flux}, \jhep{02}{2001}{030} [\hepth{0012156}].

\bibitem{Ibanez:2001nd}
L.E. Ib\'a\~nez, F.~Marchesano and R.~Rabad\'an, \emph{Getting just
  the standard model at intersecting branes}, \jhep{11}{2001}{002}
  [\hepth{0105155}].

\bibitem{850}
R.~Blumenhagen, B.~K\"ors, D.~L\"ust and T.~Ott, \emph{The standard
  model from stable intersecting brane world orbifolds},
  \npb{616}{2001}{3} [\hepth{0107138}].

\bibitem{830}
M.~Cveti\v{c}, G.~Shiu and A.M. Uranga, \emph{Three-family
  supersymmetric standard like models from intersecting brane worlds},
  \prl{87}{2001}{201801} [\hepth{0107143}].

\bibitem{Cvetic:2001kk}
M.~Cveti\v{c}, G.~Shiu and A.M. Uranga, \emph{Chiral type-II
  orientifold constructions as M-theory on $G_2$ holonomy spaces},
  \hepth{0111179}.

\bibitem{840}
R.~Blumenhagen, V.~Braun, B.~K\"ors and D.~L\"ust, \emph{Orientifolds
  of K3 and Calabi-Yau manifolds with intersecting D-branes},
  \jhep{07}{2002}{026} [\hepth{0206038}].

\bibitem{090}
R.L. Bryant and S.M. Salamon, \emph{On the construction of some
  complete metrics with exceptional holonomy}, \dmj{58}{1989}{829}.

\bibitem{080}
G.W. Gibbons, D.N. Page and C.N. Pope, \emph{Einstein metrics on
  ${S^3}$, ${R^3}$ and ${R^4}$ bundles}, \cmp{127}{1990}{529}.

\bibitem{320}
J.~Wolf, \emph{Complex homogenous contact manifolds and quaternionic
  symmetric spaces}, \newjournal{J.\ Math.\ Mech.\
  }{JOMMA}{14}{1965}{1033}.


\bibitem{990}
B.~de~Wit, B.~Kleijn and S.~Vandoren, \emph{Superconformal
  hypermultiplets}, \npb{568}{2000}{475} [\hepth{9909228}].


\bibitem{991}
B.~de~Wit, M.~Ro\v{c}ek and S.~Vandoren, \emph{Hypermultiplets,
  hyperkaehler cones and quaternion-kaehler geometry},
  \jhep{02}{2001}{039} [\hepth{0101161}].


\bibitem{992}
B.~de~Wit, M.~Ro\v{c}ek and S.~Vandoren, \emph{Gauging isometries on
  hyperkaehler cones and quaternion-Kaehler manifolds},
  \plb{511}{2001}{302} [\hepth{0104215}].


\bibitem{020}
J.F. Plebanski, \emph{A class of solutions of Einstein-Maxwell
  equations}, \ap{90}{1975}{196}.


\bibitem{010}
J.F. Plebanski and M.~Demianski, \emph{Rotating, charged and uniformly
  accelerating mass in general relativity}, \ap{98}{1976}{98}.

\bibitem{110}
B.~de~Wit and H.~Nicolai, \emph{The parallelizing $S_7$ torsion in
  gauged $N=8$ supergravity}, \npb{231}{1984}{506}.

\bibitem{100}
A.~Bilal, J.-P. Derendinger and K.~Sfetsos, \emph{(Weak) $G_2$
  holonomy from self-duality, flux and supersymmetry},
  \npb{628}{2002}{112} [\hepth{0111274}].

\bibitem{760}
S.~Chiossi and S.~Salamon, \emph{The intrinsic torsion of ${\rm SU}(3)$
  and $G_2$ structures}, \Math{DG}{0202282}.


\bibitem{691}
M.~Cveti\v{c}, G.W. Gibbons, H.~Lu and C.N. Pope, \emph{New complete
 non-compact ${\rm Spin}(7)$ manifolds}, \npb{620}{2002}{29}
 [\hepth{0103155}].

\bibitem{050}
R.~Bryant, \emph{Bochner Kaehler metrics}, \Math{DG}{0003099}.


\bibitem{040}
V.~Apostolov and P.~Gauduchon, \emph{Selfdual Einstein hermitean four
  manifolds}, \Math{DG}{0003162}.


\bibitem{030}
D.M.J. Calderbank and H.~Pedersen, \emph{Selfdual Einstein metrics
  with torus symmetry}, \Math{DG}{0105263}.


\bibitem{060}
P.-Y. Casteill, E.~Ivanov and G.~Valent, \emph{${\rm U}(1) \times
  {\rm U}(1)$ quaternionic metrics from harmonic superspace},
  \npb{627}{2002}{403} [\hepth{0110280}].

\bibitem{510}
P.-Y. Casteill, E.~Ivanov and G.~Valent, \emph{Quaternionic extension
  of the double Taub-NUT metric}, \plb{508}{2001}{354}
  [\hepth{0104078}].

\bibitem{750}
N.~Alonso-Alberca, P.~Meessen and T.~Ortin, \emph{Supersymmetry of
  topological Kerr-Newman-Taub-NUT-adS spacetimes},
  \cqg{17}{2000}{2783} [\hepth{0003071}].

\bibitem{790}
A.~Sen, \emph{Strong coupling dynamics of branes from M-theory},
\jhep{10}{1997}{002} [\hepth{9708002}].

\bibitem{450}
D.~Kramer, E.~Herlt, M.~MacCallum, and H.~Stephani, \emph{Exact
  solutions of Einstein's field equations}, Cambridge University Press
  1979.


\bibitem{120}
K.~Behrndt and G.~Dall'Agata, \emph{Vacua of $N=2$ gauged supergravity
  derived from non-homogenous quaternionic spaces},
  \npb{627}{2002}{357} [\hepth{0112136}].

\bibitem{900}
A.~Ceresole, G.~Dall'Agata, R.~Kallosh and A.~Van~Proeyen,
\emph{Hypermultiplets, domain walls and supersymmetric attractors},
\prd{64}{2001}{104006} [\hepth{0104056}].

\bibitem{500}
D.V. Alekseevsky, V.~Cortes, C.~Devchand and A.~Van~Proeyen,
  \emph{Flows on quaternionic-Kaehler and very special real
  manifolds}, \hepth{0109094}.

\bibitem{800}
R.M. Wald, \emph{General relativity}, Chicago University Press,
Chicago 1984.


\bibitem{810}
S.W. Hawking and C.J. Hunter, \emph{Gravitational entropy and global
  structure}, \prd{59}{1999}{044025} [\hepth{9808085}].

\bibitem{460}
G.W. Gibbons and S.W. Hawking, \emph{Classification of gravitational
  instanton symmetries}, \cmp{66}{1979}{291}.


\end{thebibliography}
\end{document}